\RequirePackage[2020-02-02]{latexrelease}
\documentclass[twocolumn,english,aps,prl,10pt,superscriptaddress]{revtex4}
\usepackage[T1]{fontenc}
\usepackage{amsmath,graphicx,amssymb,epsfig,babel,dsfont}
\usepackage{mathrsfs}
\usepackage{color}
\usepackage{amsbsy}

\renewcommand{\Im}{{\rm Im}}
\newcommand{\Tr}{{\rm Tr}}
\newcommand{\rd}{{\rm d}}

\newcommand{\ri}{{\rm i}}

\def\bbm[#1]{\mbox{\boldmath $#1$}}

\begin{document}

\title{Controlling local thermal states in classical many-body systems}

\author{P. Ben-Abdallah}
\email{pba@institutoptique.fr}
\affiliation{Laboratoire Charles Fabry, UMR 8501, Institut d'Optique, CNRS, Universit\'{e} Paris-Saclay, 2 Avenue Augustin Fresnel, 91127 Palaiseau Cedex, France.}
\author{A. W. Rodriguez}
\email{arod@princeton.edu}
\affiliation{Department of Electrical Engineering, Princeton University, Princeton, New Jersey 08544, USA.}

\date{\today}

\begin{abstract}
The process of thermalization in many-body systems is driven by complex interactions among sub-systems and  with the surrounding environment. Here we lay the theoretical foundations for the active control of local thermal states in arbitrary non-reciprocal systems close to their equilibrium state. In particular we describe how to (i) force some part of the system to evolve according to a prescribed law during the relaxation process (i.e. thermal targeting probem),  (ii) insulate some elements from the rest of the system or (iii) synchronize their evolution during the relaxation process. We also derive the general conditions a system must fullfill in order that some parts relax toward a minimal temperature with a minimum energetic cost or relax toward a prescribed temperature with a minimum time. Finally, we consider several representative examples in the context of systems exchanging heat radiatively.
\end{abstract}
\maketitle

Many-body heat exhange is of tremendeous importance to understand the thermalization  process  of complex networks and more generally  to understand and control their dynamic evolution. The local thermal state in these systems  is closely related to collective interactions between all elementary constituents and with the surrounding environnement. 
Many strategies have been proposed to date to actively control this evolution by way of external drives. 
Hence, by modulating some intensive quantities, such as the temperature or chemical potential, an additional flux to the primary flux induced by a temperature bias can be generated and used to control heat exchange. This shuttling effect \cite{Li_2008,Li_2009,Latella,Messina} results from the variation of the local curvature of flux with respect to these parameters. When the system displays a negative differential thermal resistance (i.e. a negative curvature of flux), this effect can contribute to inhibit the primary flux, potentially pumping heat from cold to hotter parts of the system.
A slow cycling modulation of control parameters near-topological  singularities  \cite{Li_2021,Xu} such as exceptional points  can also be used to enhance or inhibit energy exchange within a system.
Finally, the  spatiotemporal modulation of thermal properties in these systems, such as thermal conductivity or specific heat, in systems can add a convective component \cite{Torrent} to the otherwise diffusive flux. This leads to an apparent change of heat transport regime which can  be exploited to control  heat flows in solids networks at mesoscopic and macroscopic scales.

In this Letter we describe a general mathematical framework for the dynamic control of arbitrary many-body systems subject to physical constraints and discuss several problems of practical interest including local thermal targeting, thermal insulation of sub-elements and the synchronization of local thermal states during the relaxation process. We also derive the conditions that these systems must fulfill in order to accelerate the relaxation process with a minimum energetic cost and to cool some elements with a minimum time. We finally illustrate the efficiency of control on concrete situations.

To start let us consider a generic many-body system comprised of $N$ bodies interacting among themselves as well as with an external bath at temperature $T_b$. The time evolution of thermal state $\bold{T}=(T_1,...,T_N)^\top$ of this system  under temporal driving is governed by an energy balance (master) equation of general form
\begin{equation}
	C_i\frac{d{T}_i}{dt}=\underset{j\neq i}{\sum}\mathscr{P}_{j\rightarrow i}(\bold{T};T_b,t), \:\:i={1,...,N}.
	\label{Eq:dynamic1}
\end{equation}
Here $\mathscr{P}_{j\rightarrow i}$ denotes the power received by the $i^{th}$ element from the $j^{th}$ element within the system or from exchange with the external environment (i.e. thermal bath or external actuation) while $C_i$ represents its heat capacity. Close to the equilibrium state $\bold{T_{eq}}=(T_b,...,T_b)^\top$ the net power, where the symbol $\top$ set the transposition, can be linearized and expressed in terms of pairwise thermal conductance $G_{ij}=\underset{T_j \rightarrow T_i}{lim} \frac{\mathscr{P}_{j\rightarrow i}}{T_j-T_i}$. In this approximation equations (\ref{Eq:dynamic1}) can be formally recasted under the matricial form
\begin{equation}
	\frac{d\bold{T}}{dt}=\hat{\bold{G}}(t)\bold{T}+\bold{U}(t),
	\label{Eq:dynamic2}
\end{equation}
where $\hat{\bold{G}}(t)$ denotes the conductances matrix (normalized by the thermal inertia ) with $\hat{G}_{ik}=-(\underset{j\neq i}{\sum}G_{ij}+G_{ib})\delta_{ik}+(1-\delta_{ik})G_{ik}$ and $\bold{U}(t)=(u_1,...,u_N)$ is the (normalized) power vector (in $K/s$ unit) setting all interactions with the external environement, $G_{ib}$ being the conductance between the $i^{th}$ element and the bath. These interactions can be decomposed into two different contributions, the interaction of each element with the external bath and the action of external commands on these elements (i.e. $u_i=G_{ib}T_b+\mathscr{C}_i$).
It turns out that thermal state at any time reads, 
\begin{equation}
\bold{T}(t)=\mathds{R}(0,t)\bold{T}_0+\int_{0}^{t}\mathds{R}(\tau,t) \bold{U}(\tau)d\tau,\label{Nbody_solution}
\end{equation}
where $\mathds{R}$ is the resolvant of the system (\ref{Eq:dynamic2}) and $\bold{T}_0=(T_1(0),...,T_N(0))$ is the initial thermal state.
In reciprocal systems (i.e. $\hat{\bold{G}}$ is symmetric) 
\begin{equation}
\mathds{R}(s,t) =exp(\int_{s}^{t}\hat{\bold{G}}(\tau)d\tau).\label{resolvant}
\end{equation}
For systems with time-independent conductance matrix this resolvant reduces to the exponential matrix of $\hat{\bold{G}}$. The thermalization of these systems has been investigated in~\cite{Messina2,Sanders}. In non-reciprocal systems the resolvant must be calculated iteratively~\cite{SupplMat}.

\underline{Targeting of thermal state and thermal insulation:}

Starting from this general expression we can derive the conditions needed in order to locally control the temporal evolution of the system. This so called thermal targeting problem consists of finding the appropriate (external) command  in order for the system to relax toward a given targeted thermal state. The simplest targeting problem  is the temperature control of the $j^{th}$ element  of a system by using an extra power (command) $\mathscr{C}_i(t)$ ($i\neq j$) which is either injected on (i.e. $\mathscr{C}_i>0$) or extracted from (i.e. $\mathscr{C}_i<0$) the $i^{th}$ element so that the power vector reads $\bold{U}=(G_{1b}T_b,...,G_{ib}T_b+\mathscr{C}_i,...,G_{Nb}T_b)^\top$. If we denote by $T^{targ}_j$ the target temperature, then it is straighforward to show from expression (\ref{Nbody_solution}) that the optical command $\mathscr{C}^*_i$ solves the following Volterra integral equation~(see \cite{SupplMat} for details)
\begin{equation}
V_{ji}.\mathscr{C}_i(t):=\int_{0}^{t}\mathds{R}_{ji}(\tau,t) \mathscr{C}_i(\tau)d\tau=g_j(t),\label{Volterra1}
\end{equation}
with the second member
\begin{equation}
\begin{split}
g_j(t) &=T^{targ}_j(t)-\underset{k}{\sum} \mathds{R}_{jk}(0,t)T_{0k}\\
&-\underset{k}{\sum}\int_{0}^{t}\mathds{R}_{jk}(\tau,t) G_{kb}(\tau)T_b d\tau.
\end{split}
\label{gj}
\end{equation}
Moreover, the untargeted temperatures $T_{m \neq j}$ are modified by the external commands and obey the following law
\begin{equation}
T_m(t) =\hat{T}_m(t)+ \int_{0}^{t}\mathds{R}_{mi}(\tau,t) \mathscr{C}^*_i(\tau)d\tau,
\label{temp}
\end{equation}
where $\hat{T}_m$ denotes the natural (free) evolution of the temperature without external command.
It is straightforward to generalize this single targeting problem to an arbitrary number of targets.

Beside this problem we can also find the the conditions required to maintain the local thermal state associated with the $j$th element (i.e. $\dot{T}^{targ}_j(t)=0$) while the rest of the system is free to relax. The solution of this thermal insulation problem can be readily obtained by taking the time derivative of Eq. (\ref{Volterra1}) which relates the command to the local temperature. By doing so we immediately get the equation the command must satisfy the Volterra integral equation of the second kind
\begin{equation}
\int_{0}^{t}\frac{\partial \mathds{R}_{ji}(\tau,t)}{\partial t} \mathscr{C}_i(\tau)d\tau+\mathscr{C}_i(t)=\dot{g}_j(t).\label{insulate}
\end{equation}
As noted previously, the untargeted temperatures are directly impacted by the command and obey relation (\ref{temp}).

\underline{Synchronization of local thermal states:}

The second class of problem which can be considered is the synchronization of local thermal states during the relaxation process. This problem consists in finding ad hoc external commands  to synchronize the temporal evolution of different local states during the relaxation process. The simplest synchronization problem consists in finding two commands $\mathscr{C}_j(t)$ and $\mathscr{C}_k(t)$ applied on the $j^{th}$ and $k^{th}$ elements of system in order to synchronize the temperatures of two elements $m$ and $n$ along the relaxation (i.e. $T_m(t)=T_m(t)$ for $t>t^*$) . Using expression (\ref{Nbody_solution}) with  $\bold{U}=(G_{1b}T_b,...,G_{jb}T_b+\mathscr{C}_j,...,G_{kb}T_b+\mathscr{C}_k,...,G_{Nb}T_b)^\top$ and two target temperatures $T^{targ}_m(t)$ and  $T^{targ}_n(t)$  which satisfy the condition $T^{targ}_m(t)=T^{targ}_n(t)$ for $t>t^*$  it is straighforward to show that the optimal commands $\pmb{\mathscr{C}}_{jk}(t)=\left(\begin{array}{cc}
\mathscr{C}_j(t), & \mathscr{C}_k(t)\end{array}\right)^\top$ are solution of the matricial Volterra integral equation~(see \cite{SupplMat} for details)
\begin{equation}
\int_{0}^{t}\pmb{\mathds{R}}_{mn,jk}(\tau,t)\pmb{\mathscr{C}}_{jk}(\tau)d\tau=\bold{g}_{mn}(t),
\label{Volterra3}
\end{equation}
with
\begin{equation}
\pmb{\mathds{R}}_{mn,jk}(\tau,t)=\left(\begin{array}{cc}
\mathds{R}_{mj}(\tau,t) & \mathds{R}_{mk}(\tau,t)\\
\mathds{R}_{nj}(\tau,t) & \mathds{R}_{nk}(\tau,t)
\end{array}\right)
\end{equation}
and 
\begin{equation}
\bold{g}_{mn}(t)=\left(\begin{array}{cc}
T^{targ}_m(t) -\hat{T}_m(t),& T^{targ}_n(t) -\hat{T}_n(t)\end{array}\right)^\top.
\label{rmn}
\end{equation}

\underline{Active cooling with a minimum energetic cost:}

When a system is prepared at a higher temperature relative to its environment, it will cool down toward the equilibrium temperature $T_{env}$ after some general time $t^*$. According to the general expression (\ref{Nbody_solution}), the relaxation time is solution of the following equation
\begin{equation}
\bold{T}(t^*)=\mathds{R}(0,t^*)\bold{T}_0+\int_{0}^{t^*}\mathds{R}(\tau,t^*) \bold{U}(\tau)d\tau\approx \bold{T}_\infty,\label{Nbody_solution3}
\end{equation}
where $\bold{T}_\infty=(T_b,...,T_b)^\top$. This moment corresponds to the time when the global equilibrium is reached that is when all elements have the same temperature as the environment temperature. It seems obvious from this relation that the free relaxation time can in general be reduced  by extracting a sufficient amount of power from all elements having a temperature higher than the environment temperature. But such a  brute force strategy is obviously not optimal and it is strongly energy consuming.
Further below, we derive non-trivial conditions for cooling down part of the many-body system under minimum energy consumption, by either injecting or extracting power locally within the system. Formally, this problem is a linear control problem with a quadratic cost.  In its simplest form, this problem consists of solving (\ref{Eq:dynamic2}) for a power vector $\bold{U}=(G_{1b}T_b,...,G_{jb}T_b+\mathscr{C}_j,...,G_{Nb}T_b)$ subject to the quadratic cost
\begin{equation}
J(\mathscr{C}_j)=T^2_i(\bar{t})+\int_{0}^{\bar{t}} \mathscr{C}^2_j(\tau) d\tau,
\label{criteria}
\end{equation}
$\mathscr{C}_j$ being the command applied on body $j$ over a given time interval $[0,\bar{t}]$ in order to minimize the temperature of $i^{th}$ element at time $\bar{t}$ using a minimum energy consumption $E(\bar{t})=\int_{0}^{\bar{t}} \mathscr{C}^2_j(\tau) d\tau$. 
Notice that although the temperature is a positive quantity the cost function has been written, for convenience, with respect to $T^2_i$. Writing down $T_i$ explicitly in terms of in term of both the free temperature $\hat{T}_i$ and the command this function reads
\begin{equation}
J(\mathscr{C}_j)=[\hat{T}_i(\bar{t})+\int_{0}^{\bar{t}}\mathds{R}_{ij}(\tau,\bar{t})\mathscr{C}_j(\tau)d\tau]^2+\int_{0}^{\bar{t}} \mathscr{C}^2_j(\tau) d\tau.
\label{criteria2}
\end{equation}
$J$ being a strictly convex functional ($\partial^2 J/\partial \mathscr{C}^2_j=2(\beta(\bar{t})+\bar{t})>0$ where we have set $\beta(t)=\int_{0}^t\mathds{R}_{ij}(\tau,t)d\tau$) the optimal command $\mathscr{C}^*_j$ is unique. Hence, by taking the derivative of (\ref{criteria2}) with respect to $\mathscr{C}_j$ we see that the optimal command must satisfy the constraint
\begin{equation}
G[\mathscr{C}_j]=\int_{0}^{\bar{t}}\hat{K}_{ij}(\tau,\bar{t}) \mathscr{C}_j(\tau)d\tau+\hat{g}_{i}(\bar{t})=0,
\label{constraint}
\end{equation}
where we have set
\begin{equation}
\hat{K}_{ij}(\tau,\bar{t}) =\beta(\bar{t})\mathds{R}_{ij}(\tau,\bar{t})+1
\label{kernel3}
\end{equation}
and 
\begin{equation}
\hat{g}_{i}(\bar{t})=\hat{T}_i(\bar{t})\beta(\bar{t}).
\label{g_i}
\end{equation}
It follows that to minimize $E(\mathscr{C}_j)$ under the constraint (\ref{constraint}) the command must satisfy the following equation
\begin{equation}
[E+\lambda G]'(\mathscr{C}_j)=0,
\label{Beltrami}
\end{equation}
where $\lambda$ is the Lagrangian multiplier and the prime denotes the derivative of the functional $E+\lambda G$ with respect to the command. Applying the Euler-Lagrange relations for the  Lagrangian $L(\tau,\mathscr{C}_j,\dot{\mathscr{C}}_j)=\mathscr{C}^2_j+\lambda \hat{K}_{ij}(\tau,\bar{t})\mathscr{C}_j $ associated to this equation it follows that
\begin{equation}
\mathscr{C}^2_j+\lambda \hat{K}_{ij} \mathscr{C}_j =\alpha,
\label{Beltrami2}
\end{equation}
where $\alpha$ is a constant. Hence the optimal command reads
\begin{equation}
\mathscr{C}^*_j(\tau)=1/2[\lambda \hat{K}_{ij}(\tau,\bar{t}) - \sqrt{\lambda^2 \hat{K}^2_{ij}(\tau,\bar{t})+4\alpha} ],
\label{opt_command}
\end{equation}
where the constants $\lambda$ and $\alpha$ are uniquely determined from the two boundary conditions at $t=0$ and $t=\bar{t}$ for the command. After a straighforward calculation we get $\alpha=\mathscr{C}^2_j(\bar{t})$ and $\lambda=\frac{\mathscr{C}^2_j(0)-\mathscr{C}^2_j(\bar{t})}{\mathscr{C}_j(0)\hat{K}_{ij}(0,\bar{t})}$. As the minimal temperature is concerned it takes the form
\begin{equation}
T_i(\bar{t})=\hat{T}_i(\bar{t})+\int_{0}^{\bar{t}}\mathds{R}_{ij}(\tau,\bar{t})\mathscr{C}^*_j(\tau)d\tau.
\label{min_temp}
\end{equation}
\begin{figure}[bt]
		\includegraphics[angle=0,scale=0.32]{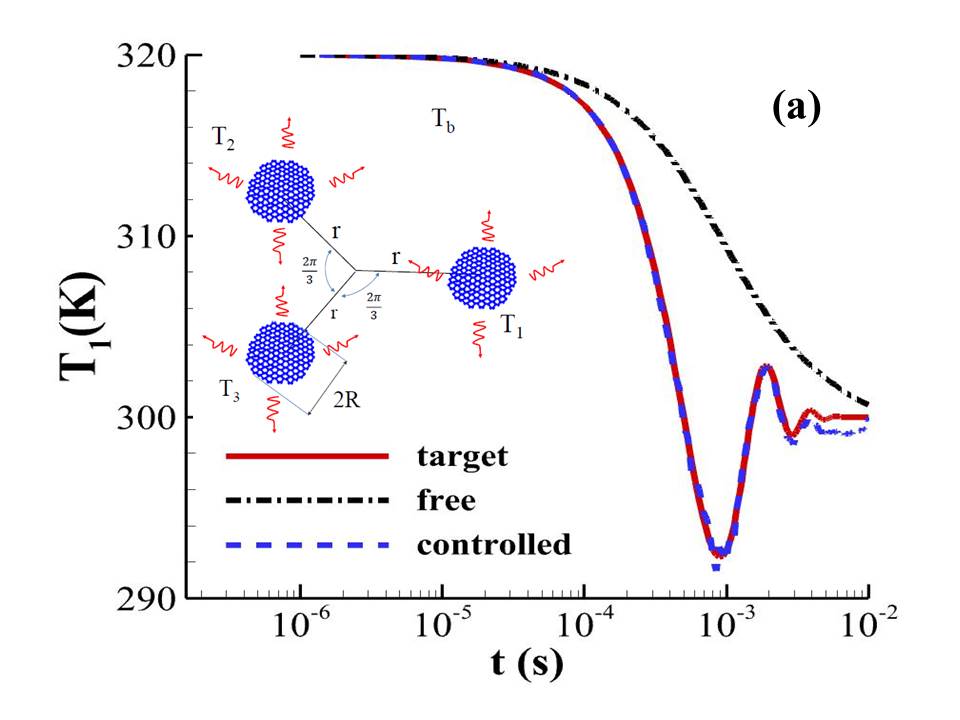}
                     \includegraphics[angle=0,scale=0.32]{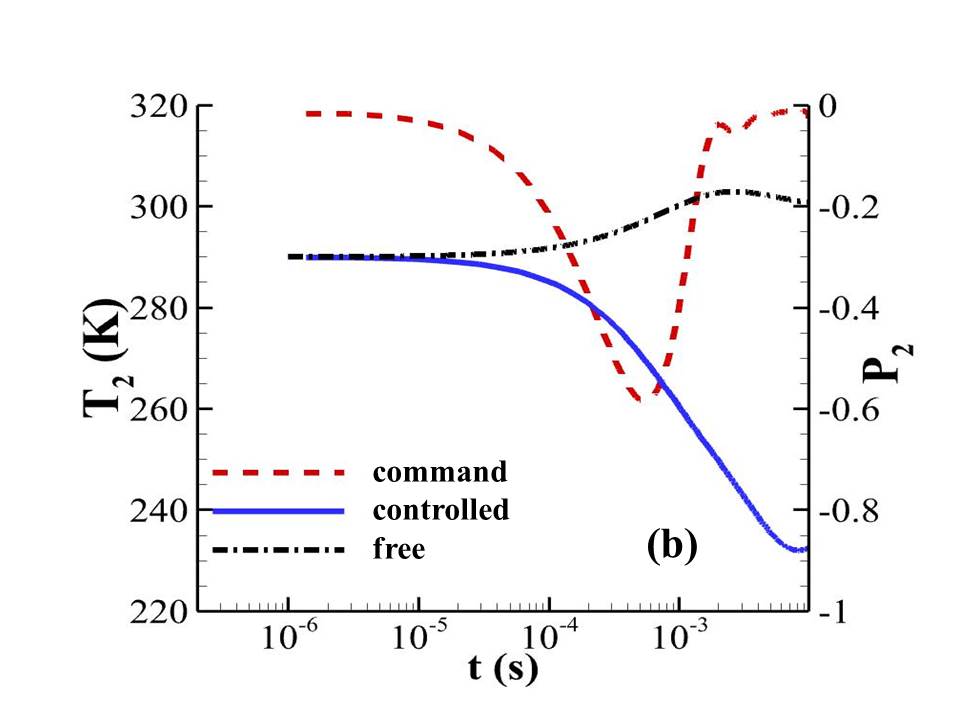}
		\caption{Targetting of the local thermal state of a three-body system consisting of graphene disks of radius $R=50\:nm$  (heat capacity $C=4.19\times 10^{-18}\: J.K^{-1}$) regularly distributed on a circle of radius $r=50\:R$ and embedded in a thermal bath at temperature $T_b=300\:K$. The disks have the same Fermi level  ($E_F=0.1\:eV$) and they are initially held at temperatures $T_1(0)=320\:K$,  $T_2(0)=290\:K$ and $T_3(0)=310\:K$. The target temperature (a) is controlled with a noisy command ($5\%$ of white noise). (b) Evolution of the temperature of controlled disk under the action of the command $\mathscr{C}_2$  and (normalized) control power $P_2=C\times\mathscr{C}_2/[G_{bb}.(T_b-T_2)]$.}
		\label{Fig:Syst}
\end{figure}
This problem can be generalized in order to identify multiple optimal command $\bold{C}^*=(\mathscr{C}^*_{a1},...,\mathscr{C}^*_{aN})$ to minimize at time $\bar{t}$  the temperature $\bold{T}=(T_{b1},...,T_{bM})^\top$ of several elements  with a minimum energy cost . In this case the cost function reads
\begin{equation}
J(\bold{C})=\parallel \bold{T}(\bar{t})\parallel_M^2+\int_{0}^{\bar{t}} \parallel \bold{C(\tau)}\parallel_N^2 d\tau.
\label{criteria3}
\end{equation}

\underline{Relaxation with a minimum of time:}

Another problem of practical importance is to consider the relaxation of the system under a minimum time constraint. In its simplest form, this implies finding a command over a bounded set $\mathscr{C}_{min}\leq \mathscr{C}_j\leq \mathscr{C}_{max}$, where $\mathscr{C}_{min}\leq 0$ and $\mathscr{C}_{max}\geq 0$, in order to minimize the relaxation time $t^*$ of $i^{th}$ element (up to the environment temperature). Formally, this problem consists in the determination of the command $\mathscr{C}_j$  which minimizes the cost $J(\mathscr{C}_j)=t^*$ so that the $i^{th}$ element reaches the surrounding temperature
\begin{equation}
 T_{env}=\hat{T}_i(t^*)+ \int_{0}^{t^*}\mathds{R}_{ij}(\tau,t^*) \mathscr{C}_j(\tau)d\tau,
\label{opt_time}
\end{equation}
$\hat{T}_i$ being the temperature followed by this element without external control. 
According to the Pontryagin maximum principle the optimal command is of the bang-bang type and reads  $\mathscr{C}^*_j(\tau)=\mathscr{C}_{min}$ if $X^{-1}_{ij}(\tau)>0$ (resp. $\mathscr{C}^*_j(\tau)=\mathscr{C}_{max}$ if $X^{-1}_{ij}(\tau)<0$) for each time $0\leq t\leq t^*$ where we have set $\bold{X}^{-1}(\tau)=exp(-\int_{0}^{\tau}\hat{\bold{G}}(s)ds)$.

We now illustrate the various external control functionals above by considering several simple examples. First, let us consider  a system composed of three graphene disks of radius $R$ placed at the vertices of an equilateral triangle  as sketched in Fig.1.  We assume this triangle is circumscribed to a circle of radius $r$ which is smaller than the  thermal wavelength of every particle and is embedded in an external bath at temperature $T_b$, such that heat exchange among the particles is dominated by near fields ~\cite{PBAEtAl2011}. In accordance with the Landauer formalism~\cite{PBAEtAl2011,RMP}, near field heat exchange between disks close to thermal equilibrium may be described by pairwise conductances,
\begin{equation}
 G_{ij}=\int_{0}^{\infty}\frac{\rd\omega}{2\pi}\frac{\partial\Theta}{\partial T}\mathcal{T}_{i,j}(\omega).
\label{Eq:Conductance}
\end{equation}
Here $\Theta(\omega,T)={\hbar\omega}/[{e^{\frac{\hbar\omega}{k_B T}}-1}]$ denotes the mean energy of a harmonic oscillator in
thermal equilibrium at temperature $T$ and $\mathcal{T}_{i,j}(\omega)$ is the transmission coefficient, at the frequency $\omega$, between two disks.
In the dipolar approximation the later reads~\cite{Cuevas,Cuevas2}
\begin{equation}
	\mathcal{T}_{ij}(\omega) = \frac{4}{3} k_0^4 \Im \Tr\biggl[\boldsymbol{\alpha}_j\mathds{G}_{ji}\frac{\boldsymbol{\alpha}_i-\boldsymbol{\alpha}_i^\dagger}{2 \ri}\mathds{G}_{ji}^{\dagger}\biggr],
\label{Eq:TransmissionCoefficient}
\end{equation}
where $k_0 = \omega/c$ with the light velocity $c$ and $\mathds{G}_{ij}$ denotes the dyadic Green tensor between the $i^{th}$ and the $j^{th}$ particle in the N-dipole system~\cite{Purcell}. Here $\boldsymbol\alpha_i$ denotes the electric polarizability tensor ~\cite{Albaladejo,de Abajo,PBA_APL2015} associated with the $i$th particle. 
Heat-exchange between the disks and the bath can be described through a correction $G_{ib}=\beta\: G_{bb}$ of the blackbody conductance $G_{bb}=4\sigma\: S \:T_b^3$ around the bath temperature $T_b$ , with a correction factor $\beta\ll 1$. Here $S$  denotes the apparent surface of disk and $\sigma$ stands for the Stefan-Boltzmann constant.
\begin{figure}[bt]
		\includegraphics[angle=0,scale=0.32]{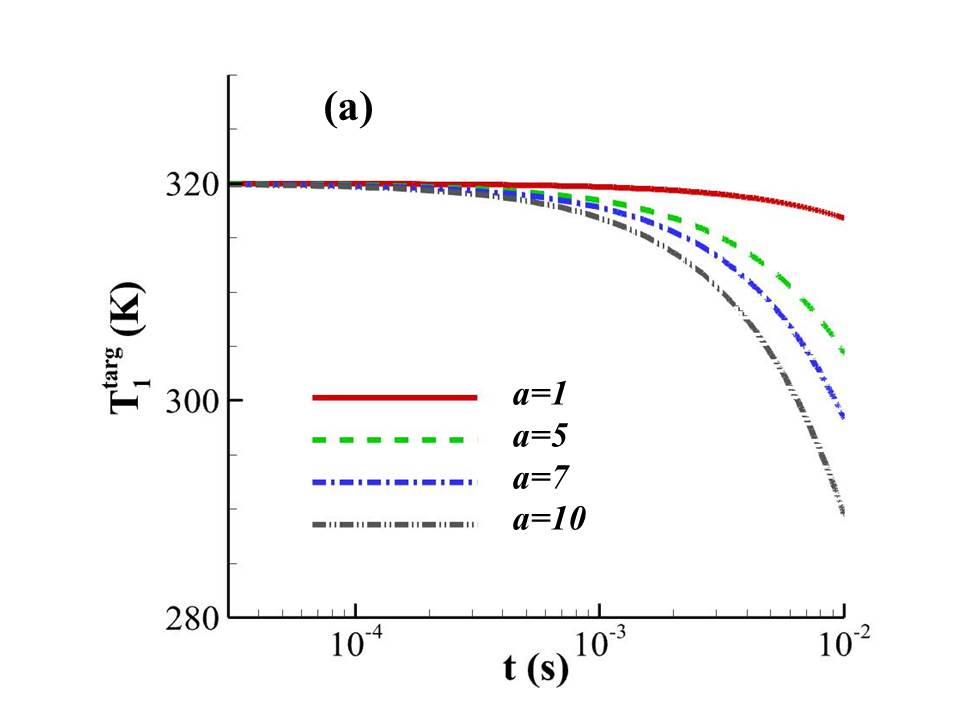}
                     \includegraphics[angle=0,scale=0.32]{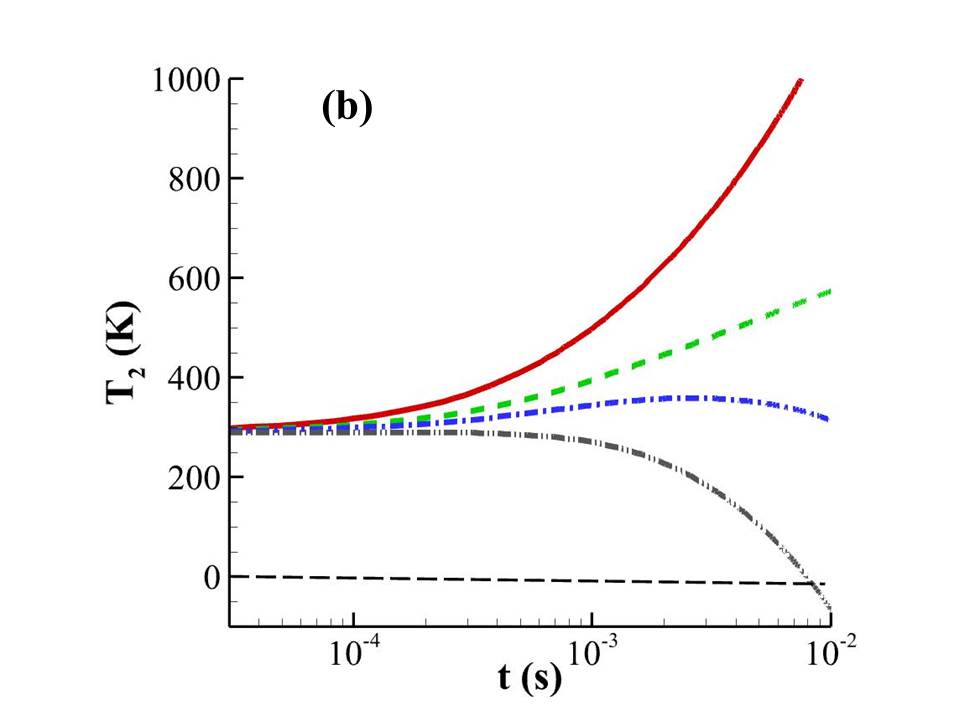}
		\caption{Reachability of a prescribed target temperature profile, $T_1^{targ}(t)=T_{1}(0)e^{-a\:t}$, for the same system of Fig. 1, under noisy command ($5\%$ of white noise) applied on disk 2. Shown are various (a) target temperature profiles correspoding to disk 1, along with the corresponding (b) effective temperature of disk 2 along the trajectory. The onset of large (even negative) temperature variations suggests some target profiles are unreachable/unphysical.}
		\label{Fig:Syst}
\end{figure}
In Fig.1(a) we plot the temporal evolution of the first disk temperature under the action of a noisy command (extracted power) applied on the second disk given an initial thermal state and we compare this evolution with its free evolution in the absence of the command. Furthermore, Fig. 1(b) shows the temperature evolution of the second disk under the external command, along with the corresponding command “cost” function. Notably, the targeted temperature is shown not only to be physically but also efficiently reachable, with both temperature and command functions always remaining physically admissible. Fig. (2) on the other hand shows that this is not always the case, as some target temperatures require either unphysical cost functions or lead to untargeted temperatures far away from equilibrium (even negative), thus invalidating the linear approximation above.
This siltuation  imposes general conditions on the commandability which must be fullfilled for a given system to reach a given target. Using the Volterra operator introduced in (\ref{Volterra1}) it is straighforward to derive from expressions (\ref{Volterra1}) and (\ref{temp}), the following commandability conditions
\begin{equation}
	T_k(t)=\hat{T}_k(t)+V_{ki}V^{-1}_{ji}g_j[T_j^{targ}(t)]\geq 0
\label{commandability}
\end{equation}
 linking the target  $T_j^{targ}$ and the local temperatures $T_k$ at  time $t$ given an initial thermal state.
\begin{figure}[bt]
		\includegraphics[angle=0,scale=0.33]{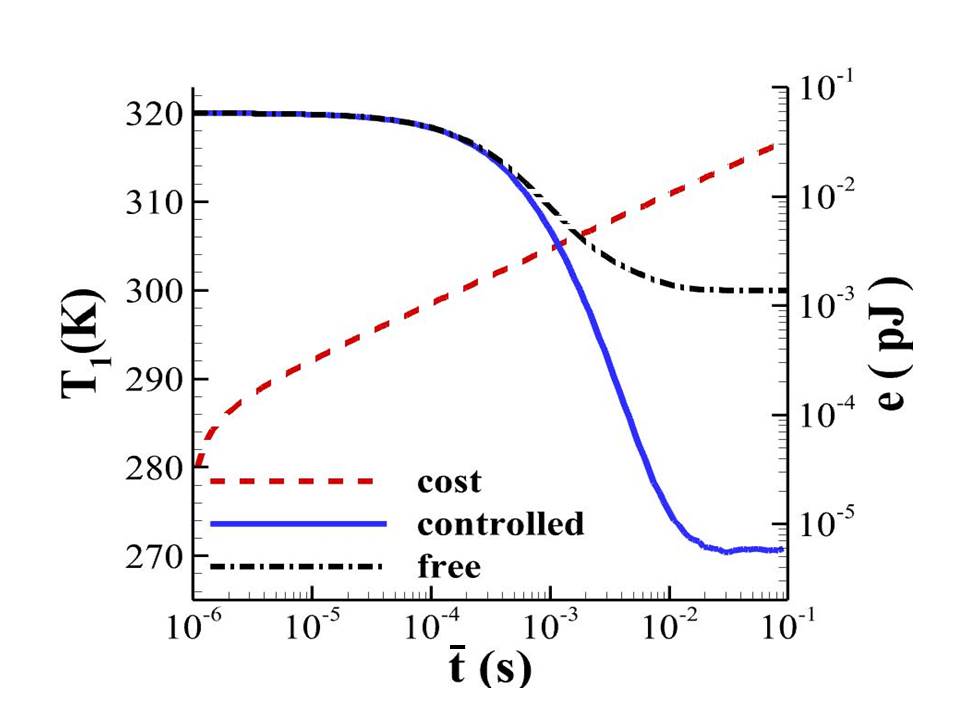}
		\caption{Optimal cooling under a minimum energy cost constraint, for the same system of Fig. 1, under noisy command ($5\%$ of white noise)  applied on disk 2. Shown on the plot are the temperature of disk 1 along with the corresponding associated energy cost $e=C\times\sqrt{\int_{0}^{\bar{t}} \mathscr{C}^{2}_j(\tau) d\tau}$ between the initial time and $\bar{t}$ where the optimal temperature is reached. The boundary conditions for the command power are set to $C\times\mathscr{C}_2(0)=4.19\:fW$ and $C\times\mathscr{C}_2(\bar{t})=21\:fW$.}
		\label{Fig:min_cost}
\end{figure}

Keeping the same system we now seek the lowest temperature the first disk can reach by controlling the temperature of disk $2$ with a minimum energetic cost. The result plotted in Fig. 3 with respect to the optimization time $\bar{t}$ given specific boundary conditions for the command (see caption of Fig.3) show that the temperature of first disk can be significantly reduced in comparison with the free relaxation process using a reasonably low energy cost. It is clear that this discrepancy can be either enlarged or reduced by an appropriate change on the boundary conditions for the command. However, in a similar way as to the targeting problem, a commandability condition analog to relation (\ref{commandability}) must be respected to keep the controlled temperatures physically admissible (close to thermal equilibrium). 
In Figs.4 we also solve the cooling problem with a minimum time on the same system using a bang-bang type control applied on the second disk. Obviously both the cooling problem with a minimum cost and the cooling problem with a minimum time strongly depends on the accessible range for the command. 

\begin{figure}[bt]
		\includegraphics[angle=0,scale=0.32]{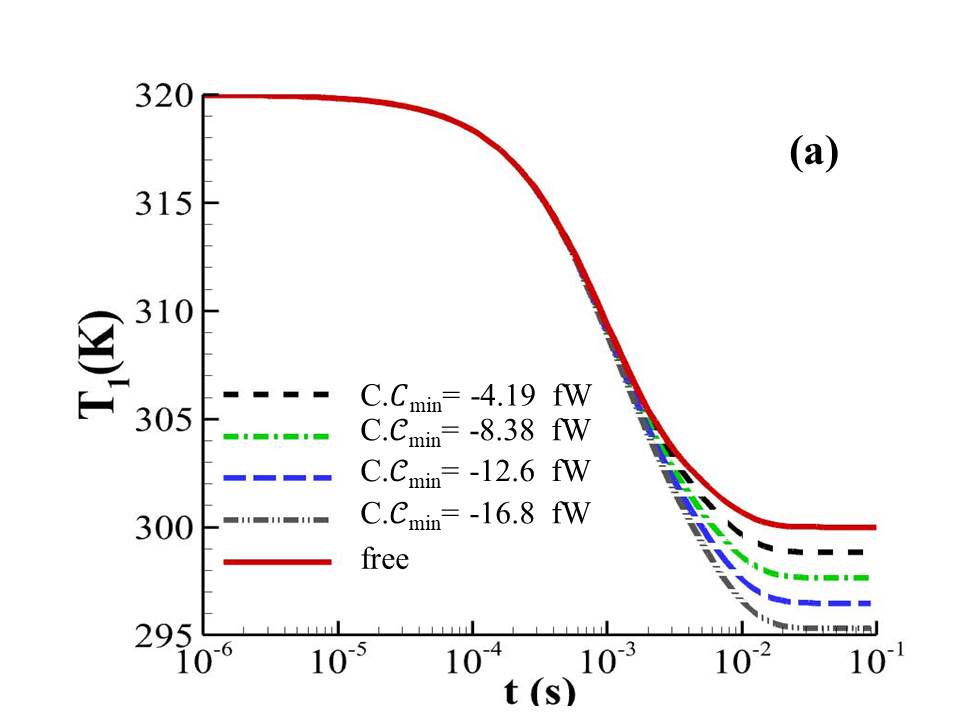}
             \includegraphics[angle=0,scale=0.32]{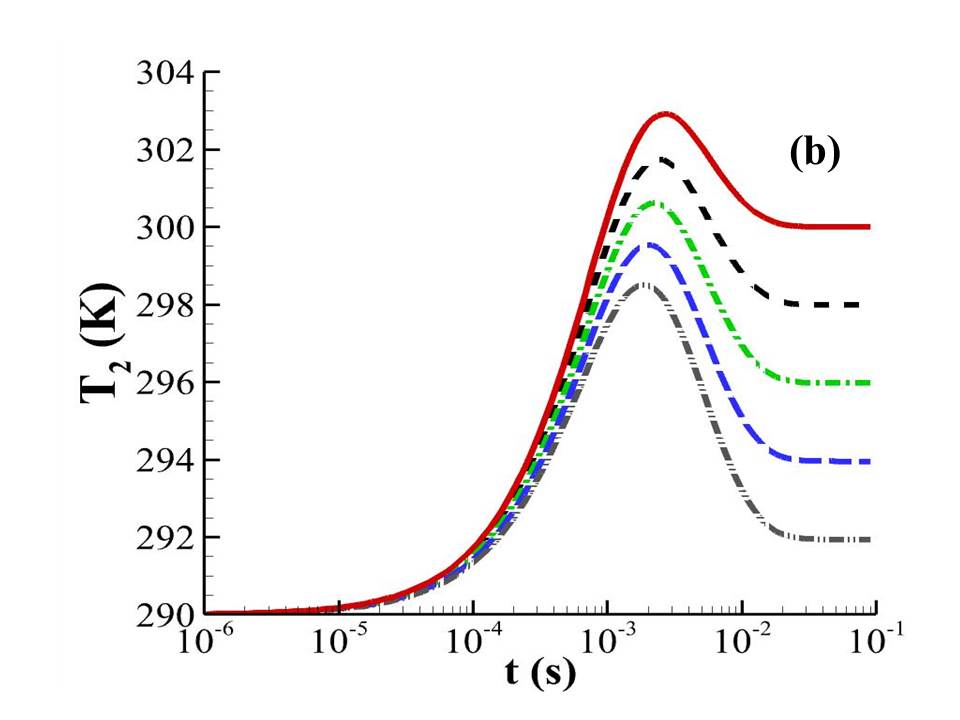}
		\caption{Optimal cooling of disk 1 with a minimum time in the same system as in Fig.1 with a bang-bang command  applied on disk $2$. The red continue curves denotes the free evolution. Temperature evolution of first disk (a) and of second disk (b) with different minima $\mathscr{C}_{min}$ for the command.}
		\label{Fig:min_time}
\end{figure}

In conclusion we laid  theoretical foundations for the active control of arbitrary many-body systems subject to physical constraints. This theoretical framework can be used to control  non-reciprocal systems with thermally dependent conductance matrices. We have demonstrated how the thermal state can be dynamically and locally manipulated using external commands showing the feasibility and versatility of various control schemes.  In particular, we have shown that  the relaxation process can be significantly accelerated or slowed down via an external drive with power requirements comparable or even smaller than the corresponding heat exchange between the elements and/or  bath. Beyond its practical interest the formalism we have introduced guarantees  convergence to globally optimal solutions and may also be exploited to derive closed form solutions potentially amenable to analytics or semi-analytical analysis. Moreover it explicitly connects this class of thermal control problems to standard problems in the area of convex optimization.  

\begin{acknowledgements}
This research was supported in part by the National Science Foundation under Grant No. PHY-1748958. P. B.-A. acknowledges the support of the French Agence Nationale de la Recherche (ANR), under grant ANR-21-CE30-0030 (NBODHEAT). 
A.R.W. acknowledges support from the National Science Foundation under the Emerging Frontiers in Research and Innovation (EFRI) programme, grant no. EFMA-1640986, the Cornell Center for Materials Research (MRSEC) through award no. DMR-1719875, and the Defense Advanced Research Projects Agency (DARPA) under grant agreements no. HR00112090011, no. HR00111820046 and no. HR0011047197.
\end{acknowledgements}


\begin{thebibliography}{99}
\bibitem{Li_2008}  N. Li, P. Hänggi and B. Li, Europhys. Lett. {\bf 84}, 40009 (2008).

\bibitem{Li_2009} N.  Li, F. Zhan, P. Hänggi and B. Li, Phys. Rev. E {\bf 80}, 011125, (2009).

\bibitem{Latella} I. Latella, R. Messina, J. M. Rubi and P. Ben-Abdallah, Phys. Rev. Lett. {\bf 121}, 023903 (2018).

\bibitem{Messina} R. Messina and P.  Ben-Abdallah, Phys. Rev. B {\bf 101}, 165435 (2020).



\bibitem{Li_2021} H. Li, L. J. Fernandez-Alcazar, F. Ellis, B. Shapiro and T. Kottos, Phys. Rev. Lett. {\bf 123}, 165901 (2019).
\bibitem{Xu} G. Xu, Y. Li, W. Li, S. Fan and C.-W. Qiu, Phys. Rev. Lett. {\bf 127}, 105901 (2021).

\bibitem{Torrent} D.Torrent, O. Poncelet and J.-C. Batsale, Phys. Rev. Lett., {\bf 120}, 125501 (2018).

\bibitem{Messina2} R. Messina, M. Tschikin, S.-A. Biehs and P. Ben-AbdallahPhys. Rev. B {\bf 88}, 104307 (2013).
\bibitem{Sanders} S. Sanders, L. Zundel, W.J.M. Kort-Kamp, D.A.R. Dalvit and A. Manjavacas, Phys. Rev. Lett. {\bf 126}, 193601 (2021).

\bibitem{SupplMat} See EPAPS Document No. [number will be inserted by publisher].

\bibitem{PBAEtAl2011} P. Ben-Abdallah, S.-A. Biehs, and K. Joulain, Phys. Rev. Lett. {\bf 107}, 114301 (2011). 
\bibitem{RMP} S.-A. Biehs, R. Messina, P.S. Venkataram, A. W. Rodriguez, J. C. Cuevas and P. Ben-Abdallah, Rev. Mod. Phys., {\bf 93}, 025009 (2021).

\bibitem{Cuevas} R. M. Abraham Ekeroth, A.  Garc\'{i}a-Mart\'{i}n, and J. C. Cuevas, Phys. Rev. B \textbf{95}, 235428 (2017).
\bibitem{Cuevas2} R. M. Abraham Ekeroth, P. Ben-Abdallah, J. C. Cuevas and A.  Garc\'{i}a-Mart\'{i}n, ACS Photonics \textbf{5} 705 (2018).

\bibitem{Purcell} E. M. Purcell and C. R. Pennypacker, Astrophys. J. {\bf 186}, 705 (1973).

\bibitem{Albaladejo} S. Albaladejo, R. G\'{o}mez-Medina, L. S. Froufe-P\'{e}rez, H. Marinchio, R. Carminati, J. F. Torrado, G. Armelles, A. Garc\'{i}a-Mart\'{i}n, and J. J. S\'{a}enz, Optics Express, \textbf{18}, 4, pp. 3556-3567 (2010).

\bibitem{de Abajo} S. Thongrattanasiri, F. H.L. Koppens, F. J. G. de Abajo, Phys. Rev. Lett., \textbf{108}, 047401 (2012).
\bibitem{PBA_APL2015} P. Ben-Abdallah, A. Belarouci, L. Frechette and S.-A. Biehs, Appl. Phys. Lett., \textbf{107}, 053109 (2015).
\end{thebibliography}
\end{document}